\documentclass[%
 reprint,
 amsmath,amssymb,
 aps,
]{revtex4-1}
\usepackage{url}
\usepackage[colorlinks=true, linkcolor=blue,urlcolor=blue,anchorcolor=blue,citecolor=blue,bookmarksnumbered]{hyperref}
\usepackage{graphicx}% Include figure files
\usepackage{dcolumn}% Align table columns on decimal point
\usepackage{bm}% bold math
\begin{document}
\title{Tripartite high-dimensional magnon-photon entanglement in $\mathcal{PT}$-symmetry broken phases of a non-Hermitian hybrid system}
\author{Jin-Xuan Han$^{1}$}\author{Jin-Lei Wu$^{1}$}\email[]{jinlei\_wu@126.com}
\author{Yan Wang$^{1}$}\author{Yan Xia$^{2}$}\author{Yong-Yuan Jiang$^{1}$}\author{Jie Song$^{1,3,4,5}$}\email[]{jsong@hit.edu.cn}
\affiliation{$^{1}$School of Physics, Harbin Institute of Technology, Harbin 150001, China\\
$^{2}$Department of Physics, Fuzhou University, Fuzhou 350002, China\\
$^{3}$Key Laboratory of Micro-Nano Optoelectronic Information System, Ministry of Industry and Information Technology, Harbin 150001, China\\
$^{4}$Key Laboratory of Micro-Optics and Photonic Technology of Heilongjiang Province, Harbin Institute of Technology, Harbin 150001, China\\
$^{5}$Collaborative Innovation Center of Extreme Optics, Shanxi University, Taiyuan, Shanxi 030006, People's Republic of China}

\begin{abstract}
Hybrid systems that combine spin ensembles and superconducting circuits provide a promising platform for implementing quantum information processing. We propose a non-Hermitian magnon-circuit-QED hybrid model consisting of two cavities and an yttrium iron garnet~(YIG) sphere placed in one of the cavities. Abundant exceptional points~(EPs), parity-time~($\mathcal{PT}$)-symmetry phases and $\mathcal{PT}$-symmetry broken phases are investigated in the parameter space. Tripartite high-dimensional entangled states can be generated steadily among modes of the magnon and photons in $\mathcal{PT}$-symmetry broken phases, corresponding to which the stable quantum coherence exists. Results show that the tripartite high-dimensional entangled state is robust against the dissipation of hybrid system, independent of a certain initial state, and insensitive to the fluctuation of magnon-photon coupling. Further, we propose to simulate the hybrid model with an equivalent $LCR$ circuit. This work may provide prospects for realizing multipartite high-dimensional entangled states in the magnon-circuit-QED hybrid system.

\end{abstract}
\maketitle
\section{Introduction}
Hybridizing two or more quantum systems can combine complementary advantages of different systems and improve the multi-task processing ability, which is the key to realizing quantum information processing~\cite{Khitrova1999,Raimond2001,Xiang2013,Aspelmeyer2014,GU20171,You2011,You2005,Buluta_2011}. Strikingly, more macroscopic objects, such as superconducting circuits possessing advantages of flexibility, scalability and tunability~\cite{Blais2004,Clarke2008,Nori2011}, are strongly coupled to electromagnetic fields, making them easy to entangle together even with shorter coherence times~\cite{DiCarlo2010,Neeley2010}. However, microscopic systems~(such as spin ensembles), naturally decoupled well from their environment and reaching relatively long coherence times~\cite{Roos2004,Balasubramanian2009}, can be integrated into a circuit by means of techniques of trapping and doping. Consequently, a hybrid system can combine the rapid operations of superconducting circuits and the long coherence time of spins.

Among possible materials of spin ensembles, a single-crystal yttrium iron garnet~(YIG) sphere has shown up recently as a promising candidate for hybrid systems, benefiting from excellent characteristics of low magnetization damping, long life, easy adjustment as well as strong coupling between magnon and photon~\cite{Mills1974,Cao2015,Zare2015,Bozhko2016}. Coherent and dissipative couplings have been identified experimentally in coupled magnon-photon hybrid systems~\cite{Soykal2010,Huebl2013,Tabuchi2014,PhysRevApplied.2.054002,Bai2015, Grigoryan2018,Harder2018,Yang2019,Xu2019,Bhoi2019,PhysRevLett.120.057202,Cao2015}. In the earlier studies, the coherent coupling between modes of photon and magnon has been proved by the anticrossing or the level repulsion of two coupled modes at/near their common resonance frequencies~\cite{Bhoi2014,Kaur2016,Maier2016,Harder2016,Castel2017}. In contrast to the anticrossing level shown in coherently coupled magnon-photon systems, dissipative coupled systems exhibit the level attraction at exceptional points~(EPs)  ~\cite{Grigoryan2018,Harder2018,Yang2019,Rao2019,Lu2021}, which opens a new avenue for exploring non-Hermitian quantum physics and parity-time~($\mathcal{PT}$) symmetry in dissipative coupled magnon-photon systems.

So far, many applications and effects have been explored based on dissipative coupled magnon-photon systems, for example nonreciprocal microwave engineering~\cite{Wang2019}, generation of the steady entangled state~\cite{Peng2020}, quantum sensing~\cite{Wolski2020}, distant magnetic
moments~\cite{Grigoryan2019,Xupeng2019} and anti-$\mathcal{PT}$ symmetry~\cite{Zhao2020}. 
Recently, Yuan $et~al$~\cite{Peng2020} reported that a high-fidelity Bell state of magnon and photon can be generated in the $\mathcal{PT}$-symmetry broken phase. Also, it has been proposed that tripartite entanglement among the deformation mode, magnetostatic mode and microwave cavity mode may be realized in a cavity magnomechanical system via magnetostrictive interaction and magnetic dipole interaction. Comparing with bipartite and tripartite binary entangled states, multipartite high-dimensional entangled states, which can enhance the violations of local realism~\cite{Kaszlikowski2000} and the security of quantum cryptography~\cite{Bo2001,Bru2002,Durkin2002,Jo2016}, have attracted much interest owing to the larger channel capacity of quantum communication and the higher efficiency of quantum information processing. To this end, high-dimensional entanglement has been not only generated theoretically in various physical platforms~\cite{Zheng_2011,Krenn2013,Su2015,Wu2017,Song2016,Wang_2020},
but also investigated experimentally in photonic systems~\cite{Kues2017,Erhard2020}, cold atoms~\cite{Parigi2015,Ding2016} and trapped ions~\cite{Senko2015}.

In this paper, we propose a non-Hermitian magnon-circuit-QED hybrid system consisting of two cavities and an YIG sphere inside one of cavities. We derive analytically an effective Hamiltonian of the hybrid magnon-photon system and its energy spectra, and then abundant EPs, $\mathcal{PT}$-symmetry phases and $\mathcal{PT}$-symmetry broken phases are investigated in the parameter space.
In $\mathcal{PT}$-symmetry broken phases, through the \textcolor{blue}{Zeeman effect} between photon and magnon modes and the electric dipole interaction between modes of photons, a tripartite high-dimensional entangled state can be generated steadily among the modes of magnon and photon, corresponding to which the stable quantum coherence exists.
Our work may facilitate potential applications of magnon-circuit-QED hybrid systems in quantum information processing, because of the following advantages and interests. First, by varying systematic parameters, there are abundant EPs, $\mathcal{PT}$-symmetry phases and $\mathcal{PT}$-symmetry broken phases in comparison with Refs.~\cite{Peng2020,Yu2020,Downing2020,Gopalakrishnan2021,Wang2021,Nair2021}. Second, the steady quantum coherence among the modes of magnon and photon exists in $\mathcal{PT}$-symmetry broken phases, with respect to which tripartite high-dimensional entangled states can be generated. However, both of quantum coherence and entanglement states appear with intense oscillations in $\mathcal{PT}$-symmetry phases, which is contrary to the universal viewpoint that the state is unstable as the $\mathcal{PT}$ symmetry is broken. This anomaly can be further understood by the competition of the evolution of non-Hermitian system and particle number conservation of the hybrid system. Finally, the fidelity of tripartite high-dimensional entangled state and the quantum coherence are robust to the dissipation of hybrid system, independent of a certain initial state, and insensitive to the fluctuation of magnon-photon coupling. By comparing with a previous study in the magnon-cavity QED hybrid system~\cite{Jie2018}, the present scheme for generating tripartite entanglement is originated from the \textcolor{blue}{Zeeman effect} between the modes of photon and magnon via non-Hermitian coupling and the electric dipole interaction between the modes of photons in the magnon-circuit-QED hybrid system. The entanglement resulted from the evolution of non-Hermitian system is not only tripartite but also high-dimensional. Therefore, the present work may provide prospects for realizing multipartite high-dimensional entangled states in the magnon-circuit-QED hybrid system and further applications in quantum information processing.
\begin{figure}[t]\centering
\includegraphics[width=\linewidth]{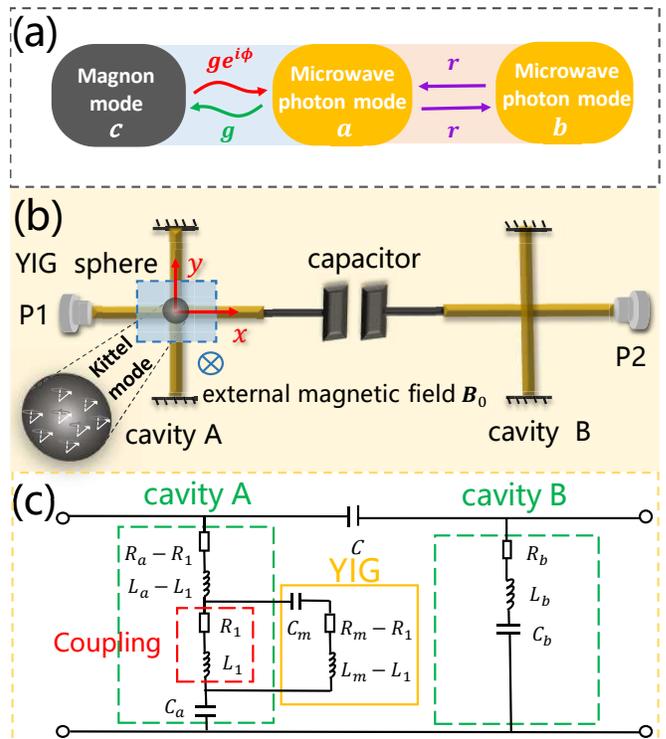}
\caption{(a)~Schematic diagram for couplings of the microwave photon mode $a$ to the microwave photon mode $b$ with strength $r$ and the magnon mode $c$ with the direct strength $g$ or feedback action strength $ge^{i\phi}$. (b)~Schematic layout of the proposed hybrid system. The design of planer cross-line microwave circuit cavity is composited by a Michelson-type microwave interferometer with two short-terminated vertical arms and two horizontal arms, in which port 1 and port 2 are connected to a vector network analyzer to enable microwave-transmission measurements~\cite{Edwards2017,satya2017}. The planar cavities $A$ and $B$ placed in the $x-y$ plane are coupled to each other by the inter-cavity capacitor. The YIG sphere magnetized to saturation, where the Kittel mode of magnon is excited by a an external magnetic field $\textbf{B}_0$ along the $z$ direction, is mounted at the center of planar microstrip-cross junction in the cavity $A$. (c)~Equivalent circuit of the coupled magnon-photon system. Circuit elements are used to model the cavity $A$~($B$) and the YIG sphere.}\label{f1}
\end{figure} 
\section{theoretical model and implementation in nanocircuit }\label{S2}

\subsection{Model and Hamiltonian}\label{S2A}
As illustrated in Fig.~\ref{f1}(a), we consider first an abstract model of a three-mode coupled magnon-photon hybrid system including two microwave photon modes $a$ and $b$ in cavities $A$ and $B$, respectively, and one magnon mode $c$ in the YIG sphere, where the photon mode $a$ is coupled to the photon~(magnon) mode $b$~($c$) via the electric~(magnetic) dipole interaction with net coupling strength $r$~($g$). In particular, there is a coherent coupling between the microwave photon mode $a$ and the magnon mode $c$ due to the Amp\'{e}re effect and the Faraday effect, corresponding to the coupling strength $g$. Meanwhile, due to the effect of Lenz's law, the microwave current in the cavity $A$ also creates a back action on the the YIG sphere to impede the magnetization of the magnon mode, which leads to a dissipative magnon-photon coupling, corresponding to the coupling strength $ge^{i\phi}$. The phase difference $\phi$ results from the competition between the coherent and dissipative couplings~\cite{Harder2018}. Such a model of the magnon-photon system can be described by a non-Hermitian interaction Hamiltonian
\begin{eqnarray}\label{e1}	
\hat{H}&=&\omega_a \hat{a}^{\dagger} \hat{a}+\omega_b \hat{b}^{\dagger} \hat{b}+\omega_c \hat{c}^{\dagger} \hat{c}+g(\hat{a}\hat{c}^{\dagger}+e^{i\phi}\hat{a}^{\dagger}\hat{c})\nonumber\\
&&+r(e^{i\theta}\hat{a}\hat{b}^{\dagger}+e^{-i\theta}\hat{a}^{\dagger}\hat{b}),
\end{eqnarray}
where $\hat{a}$~($\hat{b}$) and $\hat{c}$ are annihilation operators of the uniform precession modes for the photon mode $a$~($b$) and the magnon mode $c$. $\theta$ is a phase shift coming from the crosstalk effects between the fields produced inside the cavities $A$ and $B$. When the driving torque on the magnetization of magnon from Amp\'{e}re's law is more~(less) than the retarding torque on the magnetization of magnon from Lenz's law, $\phi=0$~($\phi=\pi$) corresponds to a purely coherent (dissipative) coupling.

\subsection{Implementation in nanocircuit}\label{S2B}
According to existing circuit-QED technologies, one can construct a magnon-circuit-QED hybrid system to implement the model proposed above via arranging planar microwave cavities $A$ and $B$ and the YIG sphere, as shown in Fig.~\ref{f1}(b). The electric dipole interaction between the cavities $A$ and $B$ can be realized by using an inter-cavity capacitor. The magnons embodied by a collective motion of a large number of spins in a ferrimagnet can be provided by an YIG sphere. At the same time, an external magnetic field $\textbf{B}_0$ along the $z$ direction magnetizes the YIG sphere to saturation and then induces the Kittel mode of magnons in the YIG sphere~\cite{Kittel1948}. The Zeeman effect is achieved by placing an YIG sphere inside the cavity $A$ so that it overlaps with the microwave magnetic field of the cavity $A$. This setup can be continuously tuned for the YIG sphere position to change the local field and the coupling effect, which can further realize dissipative coupling or coherent coupling between the YIG sphere and the cavity $A$.

In order to describe quantitatively non-Herimitian coupling in the hybrid system, both the dissipative coupling and  the coherent coupling can be realized by an $LCR$-circuit model, corresponding to the existence of the resistance-dominated coupling and the inductance-dominated coupling, respectively~\cite{Yang2019}. Equivalently, the hybrid system can be quantized into a series of circuit components of the capacitance, inductance and resistance in Fig.~\ref{f1}(c). The dissipative and coherent couplings between the YIG sphere and the cavity $A$ can be modeled by a mutual resistance $R_1$ and a mutual inductance $L_1$, respectively~\cite{Yang2019}. Also, the cavity-cavity coupling can be modeled by an inter-cavity capacitor. As long as the inter-cavity capacitance $C$ is much smaller than the  capacitance $C_a$ ($C_b$) of the cavity $A$ ($B$), photons can hop between the cavities $A$ and $B$~\cite{Schmidt2013} with the inter-cavity coupling strength $r\approx 2Z_0 C\omega_a\omega_b$, $Z_0$ being the characteristic impedance of the transmission line~\cite{Underwood2012}.
Based on the standard process quantization of an $LCR$-circuit~\cite{Blais2021}, the frequencies of the cavities $A$ and $B$ and the YIG sphere are expressed as $\omega_a=1/\sqrt{L_aC_a}$, $\omega_b=1/\sqrt{L_bC_b}$ and $\omega_c=1/\sqrt{L_mC_m}$, respectively, and the damping rates are $\gamma_a=R_a/2L_a\omega_a$, $\gamma_b=R_b/2L_b\omega_b$ and $\gamma_m=R_m/2L_m\omega_c$, respectively. 

The Hamiltonian of the hybrid system shown in Fig.~\ref{f1}(b) reads in an ideal situation as
\begin{eqnarray}\label{e2}	
\hat{H}&=&\hat{H}_{\rm{YIG}}+\hat{H}_{{A,B}}+\hat{H}_{{A-c}}+\hat{H}_{{A-B}}.
\end{eqnarray}
$\hat{H}_{\rm{YIG}}$ is the free Hamiltonian of the YIG sphere, and $\hat{H}_{{A,B}}$ is the free Hamiltonian of the cavities $A$ and $B$. $\hat{H}_{{A-c}}$~($\hat{H}_{{A-B}}$) is the interaction Hamiltonian of the cavity $A$ and the YIG sphere~(the cavity $B$), respectively. Concretely,
\begin{eqnarray}\label{e3}	
\hat{H}_{\rm{YIG}}&=&-\sum_{i}g^{*} \mu_B \textbf{B}_{0} \cdot \hat{\textbf{S}}_{i}-J \sum_{i,j} \hat{\textbf{S}}_{i} \cdot \hat{\textbf{S}}_{j}\nonumber\\
\hat{H}_{\rm{A,B}}&=&\sum_{n=a,b}\frac{1}{2} \int (\epsilon_{0} \textbf{E}_{n}^{2}+\frac{1}{\mu_0}\textbf{B}_{n}^{2})dxdydz,\nonumber\\
\hat{H}_{\rm{A-c}}&=&-\sum_{i} \hat{\textbf{S}}_{i} \cdot \textbf{H},\nonumber\\
\hat{H}_{\rm{A-B}}&=& C \int \dot{\textbf{B}}_{a} d \textbf{S}_{a} \cdot \int \dot{\textbf{B}}_{b} d \textbf{S}_{b}.
\end{eqnarray}
For $\hat{H}_{\rm{YIG}}$, $J$ is the exchange constant, $g^{*}$ the $g$-factor, $\mu_{B}$ the Bohr magneton, $\textbf{B}_{\rm{0}}$ the external magnetic field along the $z$ axis in order for the YIG sphere to be magnetized, and $\hat{\textbf{S}}_{i} \equiv (\hat{S^{x}}_i,\hat{S^{y}}_i,\hat{S^{z}}_i)$ the Heisenberg spin operator for the $i$-th site. For $\hat{H}_{\rm{A,B}}$, $\textbf{E}_{a(b)}$ and $\textbf{B}_{a(b)}$ are respective components of electric field and magnetic field in the cavity $A$~($B$), and $\epsilon_{0}$ and $\mu_0$ are vacuum permittivity and susceptibility, respectively. For $\hat{H}_{\rm{A-c}}$, $\textbf{H}$ is the corresponding magnetic field acting on the spin. For $\hat{H}_{\rm{A-B}}$, $C$ is the inter-cavity capacitance, and $\int \dot{\textbf{B}}_{n} d \textbf{S}_{n}$ is the voltage profile in the cavity $n$~($n=a,b$)~\cite{Koch2010,Nunnenkamp2011}.

The Heisenberg operators can be expressed as bosonic operators $\hat{c}_i$ and $\hat{c}^{\dagger}_i$ by using the Holstein-Primakoff transformation~\cite{Holstein1940}
\begin{eqnarray}\label{e4}
\hat{S}^{z}_{i}&=&S-\hat{c_i}^{\dagger}\hat{c_i},\nonumber\\ \hat{S}^{+}_{i}&=&\hat{c_{i}}\sqrt{2S-\hat{c_{i}}^{\dagger}\hat{c_{i}}},\nonumber\\ \hat{S}^{-}_{i}&=&\hat{c_{i}}^{\dagger}\sqrt{2S-\hat{c_{i}}^{\dagger}\hat{c_{i}}}
\end{eqnarray}
where $S$ is the total spin on each site and  $\hat{S}_{\pm} \equiv \hat{S}^{x}_{i} \pm i\hat{S}^{y}_{i}$. The bosonic operators are related to the spin-wave operators by Fourier transformation $\hat{c}_i=1/\sqrt{N}\sum_{q_{c}}e^{-i\textbf{q}_{c}\cdot \textbf{r}_{i}}\hat{c}_{q_c} ~(\hat{c}^{\dagger}_i=1/\sqrt{N}\sum_{q_{c}}e^{i\textbf{q}_{\textbf{c}}\cdot \textbf{r}_{i}}\hat{c}^\dagger_{q_{c}})$, $\hat{c}_{q_{c}}$~($\hat{c}^\dagger_{q_{c}}$) representing the annihilation (creation) operator in the spin-wave mode with wave vector $q_{c}$~\cite{kittel2005,stancil2009}. Substituting these operators into $\hat{H}_{\rm{YIG}}$, the Hamiltonian of magnetostatic modes in the YIG sphere under the static limit~\cite{stancil2009} can be written as $\hat{H}_{\rm{YIG}}=\sum_{q_c}\omega_{q_c}\hat{c}^\dagger_{q_{c}}\hat{c}_{q_c}$. Then the magnetic fields of cavities $A$ and $B$ can be quantized as $\textbf{B}_{n}=i\sum_{q_{n}}\sqrt{\omega_{q_{n}}/4V_n}\textbf{q}_{n}\times[\textbf{u}(\textbf{q}_n)\hat{n}_{q_{n}}e^{i\textbf{q}_{n}\cdot \textbf{r}}-\mathbf{u^{*}}(\textbf{q}_n)\hat{n}^\dagger_{q_{n}}e^{-i\textbf{q}_{n}\cdot\textbf{r}}]$~($n=a,b$), where $\textbf{u}(\textbf{q}_n)$ is the complex amplitude of field in the cavity, $\hat{n}_{q_{n}}$ and $\hat{n}^\dagger_{q_{n}}$ the annihilation and creation operators of the cavity at frequency $\omega_{q_n}$ with wave vector $q_{n}$, and $V_n$ the volume of the cavity. By substituting magnetic fields into $\hat{H}_{A,B}$, we obtain the quantized Hamiltonian of  microwaves $\hat{H}_{A,B}=\sum_{q_a,q_b}\omega_{q_a} \hat{a}^{\dagger}_{q_a} \hat{a}_{q_a}+\omega_{q_b} \hat{b}^{\dagger}_{q_b} \hat{b}_{q_b}$~\cite{walls2007quantum}.  

However, there are not only a coherent coupling on account of the Amp\'{e}re effect and the Faraday effect, but also a dissipative coupling on account of Lenz effect between the YIG sphere and cavity $A$~\cite{Harder2018}. In this case, the oscillating field $\textbf{H}$ acting on the local spin includes a direct action of the microwave $\textbf{h}_1$ and a reaction field of the precessing magnetization $\textbf{h}_2=\textbf{h}_1 \delta e^{i\Phi}$~\cite{Harder2018,Yang2019,Xu2019,Bhoi2019}, where $\delta$ and $\Phi$ are the relative amplitude and phase of the two fields, respectively. In the presence of both coherent and dissipative couplings, according to the standard quantization process and the rotating-wave approximation, one can recast the Hamiltonian~(\ref{e2}) into
\begin{eqnarray}\label{e5}
\hat{H}_{\rm{YIG}}&=&\sum_{q_c}\omega_{q_c}\hat{c}^\dagger_{q_{c}}\hat{c}_{q_c},\nonumber\\
\hat{H}_{\rm{A,B}}&=&\sum_{q_a,q_b}\omega_a \hat{a}^{\dagger}_{q_a} \hat{a}_{q_a}+\omega_b \hat{b}^{\dagger}_{q_b} \hat{b}_{q_b},\nonumber\\
\hat{H}_{\rm{A-c}}&=&\sum_{q_a,q_c}g_{q_a,q_c}(\hat{a}_{q_a}\hat{c}_{q_c}^{\dagger}+e^{i\phi}\hat{a}^{\dagger}_{q_a}\hat{c}_{q_c}),\nonumber\\
\hat{H}_{\rm{A-B}}&=&\sum_{q_a,q_b}r_{q_a,q_b}(e^{i\theta_{{q_a,q_b}}}\hat{a}_{q_a}\hat{b}^{\dagger}_{q_b}+e^{-i\theta_{{q_a,q_b}}}\hat{a}^{\dagger}_{q_a}\hat{b}_{q_b}),
\end{eqnarray}
where  $\phi=2\arctan[-\delta\sin{\Phi}/(1+\delta\cos{\Phi})]$ is a tunable phase~\cite{Harder2018,Yang2019,Xu2019,Bhoi2019,Peng2020}, $\theta_{q_a,q_b}$ a phase coming from the two rf-fields in the cavities $A$ and $B$ joined by microstrips with the inter-cavity capacitor, $g_{q_a,q_c}$ the magnon-photon coupling strength between the magnetostatic mode $c_{q_c}$ and the microwave cavity mode $a_{q_a}$, and $r_{q_a,q_b}$ the  photon-photon coupling strength between microwave cavity modes $a_{q_a}$ and $b_{q_b}$. The magnon-photon coupling strength via Zeeman effect and the photon-photon coupling strength via electric dipole interaction can be expressed  as~\cite{Schmidt2013,Nunnenkamp2011,tabuchi2016}  
\begin{eqnarray}\label{e6}	
g_{q_a,q_c}&=&\frac{g^{*}\mu_B\sqrt{2S}}{2} \int_V d\textbf{r}~\textbf{B}_{a} (\textbf{r}) \cdot \textbf{s}_{q_c}(\bf{r}),\nonumber\\
r_{q_a,q_c}&=&\frac{1}{2}Z_{0}C\sqrt{\omega_{q_a}(\omega_{q_b}+\Delta_{ab})}\phi_A\phi_B,
\end{eqnarray}
where $\textbf{B}_{a} (\textbf{r})$ is the strength of microwave magnetic field in the cavity $A$ at the position $\textbf{r}$ of the spin, $\textbf{s}_{q_c}(\bf{r})$ the orthonormal mode function describing the spatial profile of the amplitude and phase for the magnetostatic mode $c_{q_c}$, $Z_0$ the characteristic impedance of the transmission line, $\Delta_{ab}$ the frequency detuning between the cavities $A$ and $B$ arising from the external magnetic field $\textbf{B}_{0}$, and the $\phi_a$~($\phi_b$) the classical mode function of cavity $A$~($B$). The magnon-photon coupling is dissipative only when the YIG sphere is mounted at the center of planer microstrip-cross junction in the cavity $A$. Otherwise, it is a coherent coupling~\cite{Yang2019}.

It is supposed that the microwave magnetic field in the cavity $A$ is uniform throughout the YIG sphere, so the magnetic dipole coupling vanishes except for the uniform magnetostatic mode, i.e., the Kittel mode of magnon at frequency $\omega_c$~\cite{tabuchi2016}. The photon then only interacts strongly with the magnon around the Gamma point~(i.e, the Kittel mode) so as to match the resonant/near-resonant frequency $\omega_a$~($\omega_b$) in the cavity $A$~($B$) and $\omega_c$ in the YIG sphere~\cite{Peng2020,yuan2017,Yuan2020}. Thus the sum in Eq.~(\ref{e5}) can be removed to obtain Eq.~(\ref{e1}). It should be emphasized that the frequency $\omega_b$ can be tuned by the external magnetic field, which can further affect the cavity-cavity detuning.

\begin{figure*}
\centering
\includegraphics[width=0.86\linewidth]{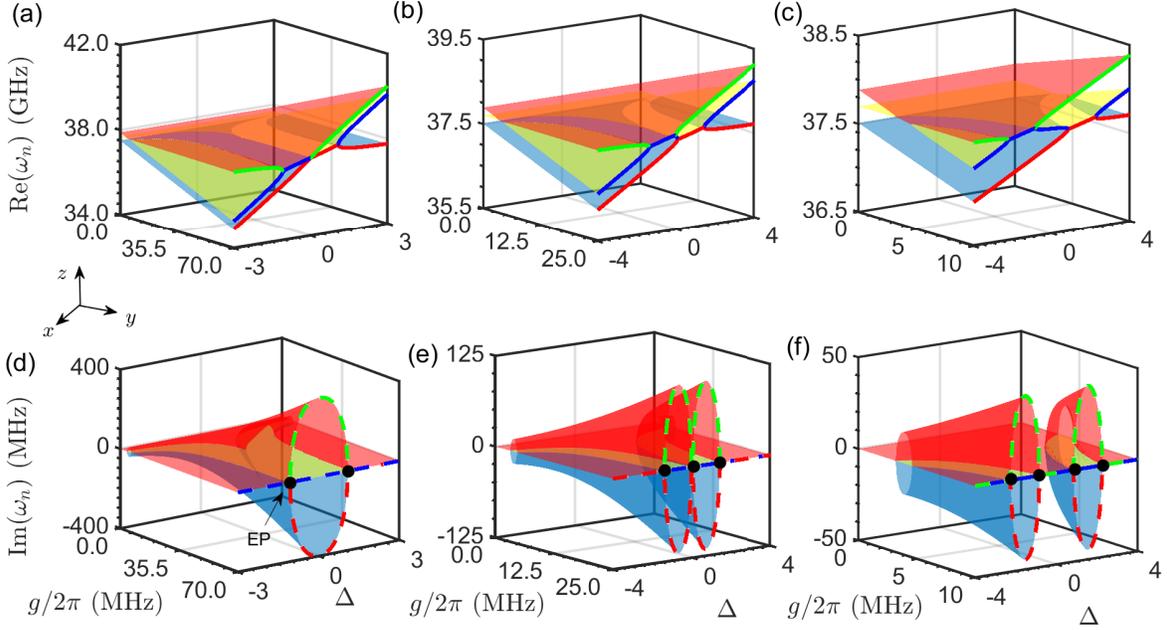}	
\caption{Energy spectrum versus the magnon-photon coupling strength $g/2\pi$ and the frequency detuning of the magnon-photon coupling $\Delta=(\omega_a-\omega_c)/2g$ by setting three decreasing magnon-photon coupling strength ranges $g/2\pi$ $\in[0,70]~$MHz in (a) and (d), $g/2\pi$ $\in[0,25]~$MHz in (b) and (e), and $g/2\pi$ $\in[0,10]~$MHz in (c) and (f), respectively. (a)-(c) Real parts of the eigenvalues. (d)-(f) Imaginary parts of the eigenvalues. $\mathcal{PT}$-symmetry broken phases in the $x-z$ plane: (a) and (d) $\Delta \in [-1.03,1.03]$ with $g/2\pi =70$~MHz; (b) and (e) $\Delta \in [-1.38,0)\cup(0,1.38]$ with $g/2\pi =25$~MHz; (c) and (f) $\Delta \in [-2.24,-0.88)\cup(0.88,2.24]$ with $g/2\pi =10$~MHz. $\mathcal{PT}$-symmetry phases in the $x-z$ plane: (a) and (d) $\Delta \in [-3,-1.03)\cup(1.03,3]$ with $g/2\pi =70$~MHz; (b) and (e) $\Delta \in [-4,-1.38)\cup(1.38,4]$ with $g/2\pi =25$~MHz; (c) and (f) $\Delta \in [-4,-2.24)\cup(-0.88,0.88)\cup(2.24,4]$ with $g/2\pi =10$~MHz. Parameters: $r/2\pi=50$~MHz, $\phi=\pi$ and $\theta=1.1\pi$.}\label{f2}
\end{figure*}

\subsection{Energy spectrum}\label{S2C}

The Hamiltonian~(\ref{e1}) is non-Hermitian, but there are still real eigenvalues~\cite{Bender1998}, because Hermicity is not a necessary condition for a physical quantum theory~\cite{Bender1998}. Importantly, under the condition of $\phi=n\pi~(n=0,1,2...)$, the Hamiltonian~(\ref{e1}) is $\mathcal{PT}$ symmetric, $[\hat{H},\mathcal{PT}] =0$, where $\mathcal{P}$ and $\mathcal{T}$ are parity inversion and time reversal operators, respectively. When the Hamiltonian~(\ref{e1}) holds real eigenvalues, the phase of the system can be regarded as the $\mathcal{PT}$-symmetry phase. However, the $\mathcal{PT}$-symmetry broken phase is characterized by complex eigenvalues~\cite{Ramy2018}. 

When the YIG sphere is mounted at the center of microstrip-cross junction in the cavity $A$, the dissipative magnon-photon coupling is absolutely dominant owing to the Lenz effect inducing a feedback microwave current on the YIG sphere and in turn impeding the excitation of the magnon, which results in the phase $\phi=\pi$~\cite{Harder2018,Yang2019}. In this situation, we can obtain analytically three complex eigenvalues $\omega_{n}$~($n=1,2,3$) of Hamiltonian~(\ref{e1})
\begin{eqnarray}\label{e7}	
\omega_1&=&\frac{\mathcal{A}}{3}-\frac{2^{1/3}(3\mathcal{B}-\mathcal{A}^2)}{3\mathcal{E}}+\frac{\mathcal{E}}{2^{1/3}}\nonumber,\\
\omega_2&=&\frac{\mathcal{A}}{3}-\frac{(1+i\sqrt{3})(3\mathcal{B}-\mathcal{A}^2)}{3\times 2^{2/3}\times \mathcal{E}^{1/3}}-\frac{(1-i\sqrt{3})\mathcal{E}^{1/3}}{6\times 2^{1/3}},\nonumber\\
\omega_3&=&\frac{\mathcal{A}}{3}-\frac{(1-i\sqrt{3})(3\mathcal{B}-\mathcal{A}^2)}{3\times 2^{2/3}\times \mathcal{E}^{1/3}}-\frac{(1+i\sqrt{3})\mathcal{E}^{1/3}}{6\times 2^{1/3}},
\end{eqnarray}
where $\mathcal{A}=\omega_a+\omega_b+\omega_c$, $\mathcal{B}=g_1^2e^{i\phi}+r^2-\omega_c\omega_a-\omega_c\omega_b-\omega_a\omega_b$, $\mathcal{C}=-\omega_c\omega_a\omega_b+\omega_cr^2+g_1^2e^{i\phi}\omega_b$, $\mathcal{D}=-\mathcal{A}^2\mathcal{B}^2+4\mathcal{B}^3-4\mathcal{A}^3\mathcal{C}+18\mathcal{ABC}+27\mathcal{C}^2$ and $\mathcal{E}=2\mathcal{A}^3-9\mathcal{AB}-27\mathcal{C}+3\sqrt{3}\sqrt{\mathcal{D}}$. We plot the real and imaginary parts of the three eigenvalues $\omega_{n}$, respectively, represented by solid and dotted lines in the $y$-$z$ plane of Fig.~\ref{f2}. Specifically, the real and imaginary parts of energy spectrum versus the magnon-photon coupling strength $g/2\pi$ and the frequency detuning of the magnon-photon coupling $\Delta=(\omega_a-\omega_c)/2g$ are also plotted in Figs.~\ref{f2}(a)-(c) and Figs.~\ref{f2}(d)-(f), respectively. We take the photon-photon coupling strength $r/2\pi=50$~MHz and  $\theta=1.1\pi$ for example and select three decreasing magnon-photon coupling strength ranges $g/2\pi$ $\in[0,70]~$MHz in (a) and (d), $g/2\pi$ $\in[0,25]$~MHz in (b) and (e), and $g/2\pi$ $\in[0,10]~$MHz in (c) and (f), respectively. It is noted that the value of coupling phase $\theta$ has no impact on the energy spectrum of the hybrid system. Obviously, the real parts of energy spectrum show two surfaces of eigenvalues merge into one hybrid surface in Figs.~\ref{f2}(a)-(c). In other words, Fig.~\ref{f2} demonstrates a typical spectrum of level attraction~($\phi=\pi$). As shown in Figs.~\ref{f2}(d)-(f), when the eigenvalues are real (imaginary), $\mathcal{PT}$-symmetry ($\mathcal{PT}$-symmetry broken) phases are appeared. The phase transitions at EPs are shown by black solid points. Depending on ranges of magnon-photon coupling strength, the regions of $\mathcal{PT}$-symmetry, $\mathcal{PT}$-symmetry broken phases and the number of EPs can be changed. 

For convenience, we take the $x-z$ plane as an example to identify the regions of $\mathcal{PT}$-symmetry and $\mathcal{PT}$-symmetry broken phases. In Figs.~\ref{f2}(a) and (d) with the magnon-photon coupling strength $g/2\pi=70$ MHz, when $\Delta \in [-3,-1.03)\cup(1.03,3]$~(real eigenvalues) and $\Delta\in [-1.03,1.03]$~(imaginary eigenvalues), there are two $\mathcal{PT}$-symmetry phases and one $\mathcal{PT}$-symmetry broken phase, respectively, with the separation of two EPs at $\Delta=\pm 1.03$. In Figs.~\ref{f2}(b) and (e) with the magnon-photon coupling strength $g/2\pi=25$~MHz, three EPs~($\Delta=-1.38,0,1.38$) divide the $\mathcal{PT}$-symmetry broken area into two adjacent parts in the region of $\Delta \in (-1.38,1.38)$. Two parts of $\mathcal{PT}$-symmetry phases exist in the region of $\Delta \in [-4,-1.38]\cup[1.38,4]$. In Figs.~\ref{f2}(c) and (f) with the magnon-photon coupling strength $g/2\pi=10$~MHz, by continuously decreasing the magnon-photon coupling strength, it is distinct that two parts of $\mathcal{PT}$-symmetry broken phases are separated further than Figs.~\ref{f2}(b) and (e). And three parts of $\mathcal{PT} $-symmetry phases are separated by four EPs at $\Delta=\pm 0.88$ and $\pm 2.24$, respectively. 
 
\section{Tripartite high-dimensional entangled states and robustness}\label{S3}
In this section, we use the model to generate the tripartite qubit entangled state and high-dimensional entangled states. Then we discuss the robustness of the entangled states and quantum coherence in the $\mathcal{PT}$-symmetry broken phases.

\subsection{Generation of tripartite qubit entangled state}\label{S3A}

We focus on the generation of tripartite qubit entangled state among the modes of magnon and photon in $\mathcal{PT}$-symmetry broken phases.
Firstly, the evolution of the hybrid system can be evaluated by using the master equation~\cite{Brody2012}
\begin{equation}\label{e8}	
\frac{\partial \hat{\rho}}{\partial t}=-i[\hat{\mathcal{H}_{1}},\hat{\rho}]-i\{\hat{\mathcal{H}_{2}},\hat{\rho}\}+2i\langle \hat{\mathcal{H}_2}\rangle \hat{\rho},
\end{equation}
where $\hat{\rho}$ is the density matrix of the system, $\hat{\mathcal{H}_{1}}$~($\hat{\mathcal{H}_{2}}$) is a Hermitian~(anti-Hermitian) operator by recasting the effective Hamiltonian as $\hat{\mathcal{H}_{1}}\equiv(\hat{H}+\hat{H}^{\dagger})/2$  ($\hat{\mathcal{H}}_{2}\equiv(\hat{H}-\hat{H}^{\dagger})/2$), $\langle \hat{\mathcal{H}}_2\rangle=\rm{tr}(\hat{\rho}\hat{\mathcal{H}}_{2})$, and the brackets $[\cdot]$ and $\{\cdot\}$ represent the commutator and anti-commutator, respectively. Specially, the resulting equation is nonlinear in the quantum state $\hat{\rho}$ by adding the third term so as to preserve $\rm{tr}(\hat{\rho})=1$.

\begin{figure}
\includegraphics[width=0.95\linewidth]{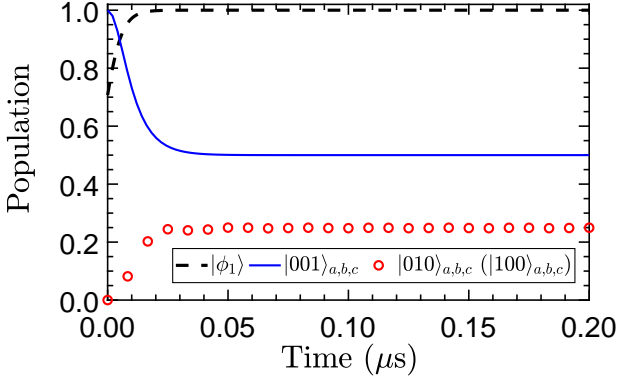}
\caption{Population evolution of the target tripartite entangled state $|\phi_1\rangle =(|\tilde{1}0\rangle_{A_1,c} +i|\tilde{0} 1\rangle_{A_1,c})/\sqrt{2}$ and its three component states, based on the initial state~$|001\rangle_{a,b,c}$ in $\mathcal{PT}$-symmetry broken phases. Parameters: $\omega_c/2\pi=6~$GHz,~$\omega_a/2\pi=\omega_b/2\pi=5.95$~GHz,~$g/2\pi=6$~MHz, $r/2\pi=50$~MHz and $\theta=1.1\pi$.}\label{f3}
\end{figure}

\begin{figure}
\includegraphics[width=0.88\linewidth]{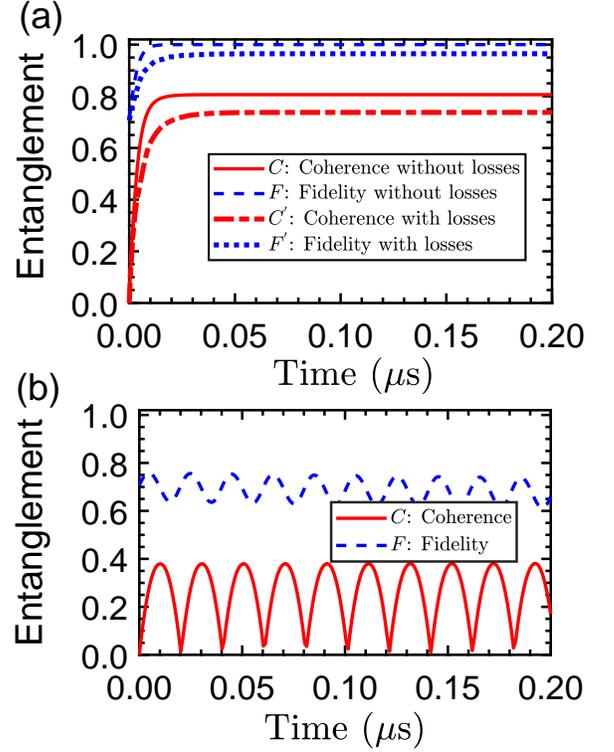}
\caption{Time dependence of the collective coherence of the system and the fidelity of $|\phi_1\rangle =(|\tilde{1}0\rangle_{A_1,c} +i|\tilde{0} 1\rangle_{A_1,c})/\sqrt{2}$, based on the initial state~$|001\rangle_{a,b,c}$ (a) in the $\mathcal{PT}$-symmetry broken phase and (b) in the $\mathcal{PT}$-symmetry phase. Parameters: (a) $\omega_c/2\pi=6~$GHz,~$\omega_a/2\pi=\omega_b/2\pi=5.95$~GHz,~$g/2\pi=6$~MHz,~$r/2\pi=50$~MHz and $\kappa_a/2\pi=\kappa_b/2\pi=\gamma_m/2\pi=0.1g$; (b) $\omega_c/2\pi=6~$GHz,~$\omega_a/2\pi=\omega_b/2\pi=6$~GHz,~$g/2\pi=6$~MHz, $r/2\pi=50$~MHz and $\theta=1.1\pi$.}\label{f4}
\end{figure}

By solving the evolution of density matrix $\rho$ governed by Eq.~(\ref{e6}), an initial pure state $|\phi_0\rangle=|001\rangle_{a,b,c}$ with a mean particle number $N=\langle \hat{a}^{\dagger}\hat{a}+\hat{b}^{\dagger}\hat{b}+\hat{c}^{\dagger}\hat{c}\rangle =1$ is taken as an example. Then we introduce nonlocal modes
\begin{eqnarray}\label{e9}	
\hat{A}_{1,2}=\frac{\hat{a} \pm \hat{b}e^{-i\theta}}{\sqrt{2}}.
\end{eqnarray}
And assume that $\omega_a=\omega_b=\omega$ and the large detuning condition $|\omega-\omega_c-r|\gg|g|$. After transforming into the interaction picture of the nonlocal modes, the interaction Hamiltonian reads as 
\begin{equation}\label{e10}
\hat{H}_{\rm{eff}}=g_{\rm{eff}}(\hat{A}_{1}^{\dagger}\hat{c}+\hat{A}_{1}\hat{c}^{\dagger}e^{i\phi}),
\end{equation}
where $\hat{A}_{1}=(\hat{a}+\hat{b}e^{-i\theta})/\sqrt{2}$ and $g_{\rm{eff}}=g/\sqrt{2}$.  Accordinlgly, the hybrid system evolves in the finite subspace $\{|\tilde{0} 1\rangle_{A_1,c},|\tilde{1} 0\rangle_{A_1,c} \}$, where $|\tilde{0}\rangle_{A_1}$ and $|\tilde{1}\rangle_{A_1}$ are Fock states of the mode $A_1$, represented by $|00\rangle_{a,b}$ and $(|10\rangle_{a,b}+e^{i\theta}|01\rangle_{a,b})/\sqrt{2}$, respectively, on the basis of $\{|0\rangle_{a}, |1\rangle_{a}, |0\rangle_{b}, |1\rangle_{b}\}$.  

By solving the eigenequation $\hat{H}|\phi_m\rangle=\omega_m|\phi_m\rangle$, eignstates of the hybrid system are obtained 
\begin{equation}\label{e11}
|\phi_m\rangle=\cos{\theta_{m}}|\tilde{1}0\rangle_{A_1,c}+e^{i\psi_m}\sin{\theta_m}|\tilde{0}1\rangle_{A_1,c},
\end{equation}
where  $\theta_m$ and $\psi_m$ are related with $e^{i\psi_m}\tan {\theta_m}=(\omega_k-\omega)/g~(k=1,2,3...)$. If the initial state is $\hat{\rho}_0=|\tilde{0} 1\rangle_{A_1,c}\langle \tilde{0}1|$, the time-dependent density matrix can be further written as~\cite{Brody2012}
\begin{equation}\label{e12}
\hat\rho=\frac{\sum_{k,j}p_{k,j}e^{-i\omega_{kj}t}|\phi_k\rangle\langle\phi_j|}{\sum_{k,j}p_{k,j}e^{-i\omega_{kj}t}{\rm tr}(|\phi_k\rangle\langle\phi_j|)}, 
\end{equation}
where $\omega_{k,j}=\omega_k-\omega_{j}^{*}$ and $p_{k,j}$ are expansion coefficients. As $t\rightarrow \infty$, the steady density matrix is $\hat{\rho}(\infty)=|\phi_1\rangle \langle \phi_1|$. On the one hand, the steady state is $|\phi_1\rangle =(|\tilde{1}0\rangle_{A_1,c} +i|\tilde{0} 1\rangle_{A_1,c})/\sqrt{2}$, which is a Bell state in the general form on the basis of $\{|\tilde{0} 1\rangle_{A_1,c},|\tilde{1} 0\rangle_{A_1,c} \}$. On the other hand, it is a tripartite entangled state with $|\phi_1\rangle=|100\rangle_{a,b,c}/2+e^{i\theta}|010\rangle_{a,b,c}/2+i|001\rangle_{a,b,c}/\sqrt{2}$ on the basis of $\{|0\rangle_{a}, |1\rangle_{a}, |0\rangle_{b}, |1\rangle_{b}, |0\rangle_{c}, |1\rangle_{c}\}$.

In Fig.~\ref{f3}, we plot numerically population evolution of the system, including the ideal entangled state and other evolution states in $\mathcal{PT}$-symmetry broken phases. By satisfying the large detuning condition, we set $\omega_c/2\pi=6$~GHz, $\omega_a/2\pi=\omega_b/2\pi=5.95$~GHz, $g/2\pi=6$~MHz and $r/2\pi=50$~MHz that is accessible in ~\cite{Underwood2012,Schmidt2013,Huebl2013,Tabuchi2014,Yang2019,Xu2019,Lachance2019}. The fidelity is formulated as $F(t)={\rm{tr}}\sqrt{\langle \phi_1|\hat{\rho}(t)|\phi_1\rangle}$, where $|\phi_1\rangle$ and $\hat{\rho}(t)$ are the target state and the time-dependent density matrix of the system by solving Eq.~(\ref{e8}). It is identified that the population of $|010\rangle_{a,b,c}$ and $|100\rangle_{a,b,c}$ has the exactly identical evolution with population $0.25$ while the population of the initial state $|001\rangle_{a,b,c}$ becomes $0.5$ at the time $T=0.2~\mu$s. Evidently, the population of the target state $|\phi_1\rangle$ evolves from about 0.7 to 1, signifying the successful creation of a steady tripartite entangled state in $\mathcal{PT}$-symmetry broken phases. 

In fact, in a tripartite system, quantum coherence may exist due to the collective participation of several subsystems, or can be attributed to coherence located within the subsystems. Therefore, the magnon-photon entanglement can be quantified through the collective coherence, which are given by expression~\cite{Chandrashekar2019}
\begin{eqnarray}\label{e13}	
C&=&\sqrt{S(\frac{\hat{\rho}+\hat{\rho}_{\pi}}{2})-\frac{S(\hat{\rho})+S(\hat{\rho}_{\pi})}{2}},
\end{eqnarray}
in which $S$ is the von Neumann entropy, $\hat{\rho}$ the density matrix of the hybrid system, and the closest product state $\hat{\rho}_{\pi}\equiv \hat{\rho}_{\rm{min}}=\hat{\rho}_a \otimes \hat{\rho}_b \otimes \hat{\rho}_c$. Figures \ref{f4}(a) and (b) show the time evolution of fidelity~(thin dotted line) of $|\phi_1\rangle$ and the collective coherence~(thin solid line) without losses in the $\mathcal{PT}$-symmetry broken phase and the $\mathcal{PT}$-symmetry phase, respectively. Apparently, in the $\mathcal{PT}$-symmetry broken phase with complex eigenvalues $\omega_{n}$, $F$ and $C$ approach one and 0.806, respectively, when the system without losses evolves to the tripartite entangled state $|\phi_1\rangle$. Nevertheless, the system enters into the $\mathcal{PT}$-symmetry phase as the eigenvalues $\omega_{n}$ are real. The hybrid system has no gain modes, which results in the unstable dynamics of collective coherence and fidelity. Thus, the steady tripartite entangled state and the collective coherence can be steady in the $\mathcal{PT}$-symmetry broken phase but oscillate in the $\mathcal{PT}$-symmetry phase.

\begin{figure}[htb]\centering
\centering
\includegraphics[width=0.88\linewidth]{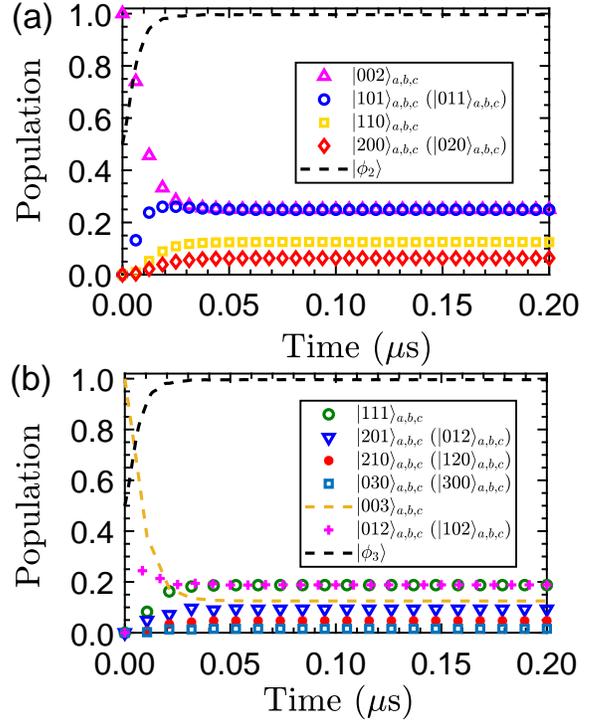}
\caption{Population evolution of the target tripartite high-dimensional entangled states and their component states for (a)~$|\phi_2\rangle=|\tilde{2} 0\rangle_{A_1,c}/2-i|\tilde{1} 1\rangle_{A_1,c}/\sqrt{2}-|\tilde{0} 2\rangle_{A_1,c}/2$ based on the initial state $|002\rangle_{a,b,c}$ and (b) $|\phi_3\rangle=i\sqrt{2}|\tilde{0}3\rangle_{A_1,c}/4+\sqrt{6}|\tilde{1}2\rangle_{A_1,c}/4-i\sqrt{6}|\tilde{2}1\rangle_{A_1,c}/4-\sqrt{2}|\tilde{3}0\rangle_{A_1,c}/4$ based on the initial state $|003\rangle_{a,b,c}$. Parameters: $\omega_c/2\pi=6~$GHz,~$\omega_a/2\pi=\omega_b/2\pi=5.95$~GHz,~$g/2\pi=6$~MHz, $r/2\pi=50$~MHz and $\theta=1.1\pi$. }\label{f5}
\end{figure}

Actually, this anomaly can be further understood by the competition of the evolution of non-Hermitian system and the particle number conservation in the hybrid system. When the hybrid system lies in the $\mathcal{PT}$-symmetry broken phase, the evolution of the system guarantees not only the process of gain and loss but also the particle conserving process. The coexistence of two processes will render the initial state to evolve in the steady target entangled state. As for the $\mathcal{PT}$-symmetry phase, evolution of the hybrid system does not involve gain and loss, and hence can not render the system to a steady state. The oscillation phenomenon is analogous to the unitary evolution of traditional Hermitian systems with a real beam-splitter type interaction through the particle conserving process~\cite{yuan2021quantum}.

\begin{figure}[htb]\centering
\centering
\includegraphics[width=0.9\linewidth]{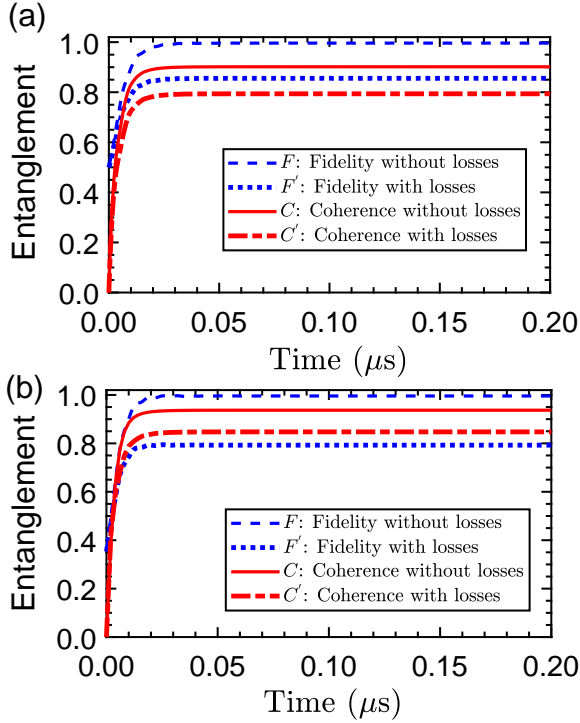}
\caption{Time dependence of the collective coherence and fidelity of tripartite high-dimensional entangled states (a) $|\phi_2\rangle=|\tilde{2} 0\rangle_{A_1,c}/2-i|\tilde{1} 1\rangle_{A_1,c}/\sqrt{2}-|\tilde{0} 2\rangle_{A_1,c}/2$ based on the initial state $|002\rangle_{a,b,c}$  and (b) $|\phi_3\rangle=i\sqrt{2}|\tilde{0}3\rangle_{A_1,c}/4+\sqrt{6}|\tilde{1}2\rangle_{A_1,c}/4-i\sqrt{6}|\tilde{2}1\rangle_{A_1,c}/4-\sqrt{2}|\tilde{3}0\rangle_{A_1,c}/4$ based on the initial state $|003\rangle_{a,b,c}$ in the $\mathcal{PT}$-symmetry broken phase. Parameters:  $\omega_c/2\pi=6~$GHz,~$\omega_a/2\pi=\omega_b/2\pi=5.95$~GHz,~$g/2\pi=6$~MHz,~$r/2\pi=50$~MHz, $\theta=1.1\pi$ and $\kappa_a=\kappa_b=\gamma_m=0.1g$.}\label{f6}
\end{figure}

\subsection{Generation of tripartite high-dimensional entangled states}
In the following, we consider the hybrid system evolving in the Hilbert subspace of $N>1$, when the system is under the $\mathcal{PT}$-symmetry broken phase.

Firstly, we take an example of an initial pure state $|\psi_2\rangle=|\tilde{0} 2\rangle_{A1,c}$ with a
mean particle number $N=2$. By solving the master equation (\ref{e6}), a steady tripartite three-dimensional entangled state can be generated by satisfying the large detuning condition, which is represented as
\begin{equation}\label{e14}
|\phi_2\rangle=\frac{1}{2}|\tilde{2} 0\rangle_{A_1,c}-\frac{i}{\sqrt{2}}|\tilde{1} 1\rangle_{A_1,c}-\frac{1}{2}|\tilde{0} 2\rangle_{A_1,c},
\end{equation}
where $|\tilde{2}\rangle_{A_1}$ can be represented by $(|20\rangle_{a,b} +e^{i\theta}\sqrt{2}|11\rangle_{a,b}+e^{2i\theta}|02\rangle_{a,b})/2$,  on the basis of $\{|0\rangle_{a}, |1\rangle_{a}, |2\rangle_{a}, |0\rangle_{b}, |1\rangle_{b}, |2\rangle_{b}\}$. In Fig.~\ref{f5}(a), with the same parameters as in Fig.~\ref{f3}, we numerically plot the time evolution of populations for the target entangled state $|\phi_2\rangle$, the initial state $|\psi_2\rangle$, and other evolution states. The populations of $|101\rangle_{a,b,c}$ and $|011\rangle_{a,b,c}$~($|020\rangle_{a,b,c}$ and $|200\rangle_{a,b,c}$) have the identical trend reaching 0.25~(0.06). The population of initial state $|002\rangle_{a,b,c}$ evolve from 1 to 0.25, and the population of $|110\rangle_{a,b,c}$ is close to 0.16 finally. Obviously, the population of $|\phi_2\rangle$ reaches nearly 1 and remains steady at the end of evolution, which indicates the successful creation of the tripartite high-dimensional entangled state.

Next, by setting initially a mean particle number $N=3$ and choosing an initial state as $|\psi_3\rangle=|\tilde{0} 3\rangle_{A1,c}$, we obtain a tripartite four-dimensional entangled state
\begin{equation}\label{e15}
|\phi_3\rangle=\frac{\sqrt{2}}{4}(\sqrt{3}|\tilde{1}2\rangle_{A_1,c}-i\sqrt{3}|\tilde{2}1\rangle_{A_1,c}-|\tilde{3}0\rangle_{A_1,c}+i|\tilde{0}3\rangle_{A_1,c}),
\end{equation} 
where $|\tilde{3}\rangle_{A_1}$ can be represented by $(e^{2i\theta}\sqrt{6}|12\rangle_{a,b}+e^{i\theta}\sqrt{6}|21\rangle_{a,b}+\sqrt{2}|30\rangle_{a,b}+e^{3i\theta}\sqrt{2}|03\rangle_{a,b})/4$,  on the basis of $\{|0\rangle_{a}, |1\rangle_{a}, |2\rangle_{a}, |3\rangle_{a}, |0\rangle_{b}, |1\rangle_{b}, |2\rangle_{b}, |3\rangle_{b}\}$. Also, the population of the initial state, the target state and other evolution states are exhibited in Fig.~\ref{f5}(b) with the same parameters as Fig.~\ref{f5}(a). The fidelity of $|\phi_3\rangle$ reaches unity and remains stable at the end of evolution. Thus, the result reveals that the tripartite high-dimensional entangled state can be generated in this scheme. Figures \ref{f6}(a) and (b) show the time evolution of the collective coherence~(thin solid line) and fidelity~(thin dotted line) with $|\phi_2\rangle$ and $|\phi_3\rangle$ in the $\mathcal{PT}$-symmetry broken phase, respectively. As expected, the fidelity of $|\phi_2\rangle$ and $|\phi_3\rangle$ can be both achieved by unity in the $\mathcal{PT}$-symmetry broken phase, when not considering losses. The final collective coherence of $|\phi_2\rangle$ and $|\phi_3\rangle$ are 0.902 and 0.937, respectively, and keep steady. The tripartite entangled state has the larger collective coherence when the initial state is of a larger mean particle number.

\begin{figure}
\centering
\includegraphics[width=\linewidth]{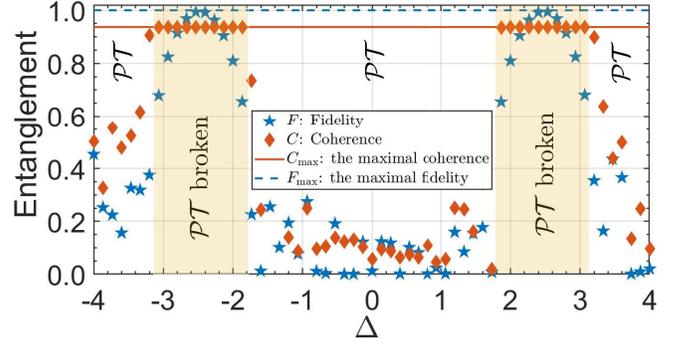}
\caption{Collective coherence and the fidelity of the target tripartite high-dimensional entangled state $|\phi_3\rangle$ as functions of the frequency detuning of the magnon-photon coupling $\Delta$. Parameters: $\omega_c/2\pi=6~$GHz,~$\omega_a/2\pi=\omega_b/2\pi=5.95$~GHz,~$g/2\pi=6$~MHz, $r/2\pi=50$~MHz, $\theta=1.1\pi$ and $T_{\rm{total}}=0.2~\mu$s.}\label{f7}
\end{figure}

In order to represent more intuitively the relation among the frequency detuning of the magnon-photon coupling, the collective coherence, and the fidelity of tripartite high-dimensional entangled state, a full phase diagram of the system is shown in Fig.~\ref{f7}. In $\mathcal{PT}$-symmetry broken phases, the collective coherence remains the maximum and steady value, independent of the magnitude of detuning. Meanwhile, the fidelity becomes unity under the condition of $\Delta\approx\pm2.5$. We find that the collective coherence and the fidelity are of oscillations and cannot reach the maximal value in the $\mathcal{PT}$-symmetry phases. Therefore, the EPs at $\Delta=\pm1.9$ and $\pm3.1$ play an important role of critical point whether the target tripartite high-dimensional entanglement can be generated steadily.

\begin{figure}
\centering
\includegraphics[width=1\linewidth]{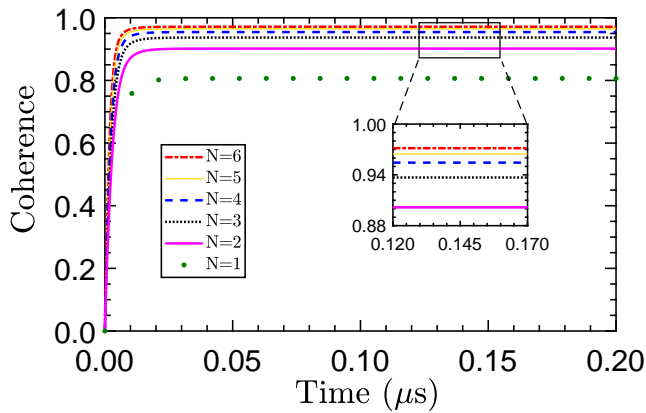}	
\caption{Collective coherence as the function of evolution time with the total number of particles from $N=1$ to $N=6$ based on the initial state $|00N\rangle_{a,b,c}$. Parameters:  $\omega_c/2\pi=6~$GHz,~$\omega_a/2\pi=\omega_b/2\pi=5.95$~GHz,~$g/2\pi=6$~MHz, $r/2\pi=50$~MHz and $\theta=1.1\pi$.}\label{f8}
\end{figure}

In Fig.~\ref{f8}, we numerically work out the time evolution of the collective coherence with the total number of particles from 1 to 6. Apparently, the collective coherence has the lager value with the increase of $N$. Nevertheless, when $N=2,3,4,5,6$, the collective coherence increases slowly and lies in the position between 0.937 and 0.972. By magnifying the time range $T\in[0.12,0.17]~\mu$s, tripratite high-dimension entangled states can be obtained with the collective coherence of 0.902, 0.937, 0.955, 0.965 and 0.972, respectively, when $N=2,3,4,5,6$. As for the maximal tripartite high-dimensional entangled states $|\Psi_n\rangle = \frac{1}{\sqrt{N+1}}(\sum^{N}_{k=0}|k,k,k\rangle_{a,b,c})$, their collective coherence can reach 0.941, 0.970, 0.983, 0.989 and 0.993, respectively, corresponding to $N=2,3,4,5,6$, calculated by Eq.~(\ref{e13}). In fact, tripartite high-dimensional entangled states proposed here are not the standard ones $|\Psi_n\rangle$, and hold collective coherence slightly less than the maximal ones for the integer $N\in [2,6]$. However, tripartite high-dimensional entangled states $|\phi_n\rangle$ have also enough capacity to complete tasks of quantum information. Besides, the tripartite high-dimensional entangled state $|\phi_n\rangle$ can still collapse possibly to entanglement states, such as $|\tilde{1}\rangle_{A_1}$, $|\tilde{2}\rangle_{A_1}$ and $|\tilde{3}\rangle_{A_1}$, when quantum measurements are introduced.

\subsection{Robustness of entanglement}
In this subsection, we discuss the robustness of the tripartite high-dimensional entangled state against environment noises. Three dominant influence aspects are considered: (i) losses of cavity $A$, cavity $B$ and the magnon with decay rates $\kappa_a$, $\kappa_b$ and $\gamma_m$, respectively; (ii) different initial states; (iii) unexpected magnon-photon coupling $g$. 

When the hybrid system with $\mathcal{PT}$-symmetry broken phases is in an ideal environment, the result shows that the collective coherence and the fidelity of tripartite entangled states $|\phi_1\rangle$, $|\phi_2\rangle$ and $|\phi_3\rangle$ remain stable in Figs.~\ref{f4}(a), \ref{f6}(a) and \ref{f6}(b). In order to give a quantitative illustration of the effects of losses of cavity modes and the magnon mode, the dynamics of the lossy system is governed by adding Lindblad operators of cavity modes and the magnon mode into the master equation 
\begin{eqnarray}\label{e16}
\frac{\partial \hat{\rho}}{\partial t}&=&-i[\hat{\mathcal{H}}_{1},\hat{\rho}]-i\{\hat{\mathcal{H}}_{2},\hat{\rho}\}+2i\langle \hat{\mathcal{H}}_2\rangle \hat{\rho}+\gamma_mD[\hat{c}]\nonumber\\
&&+\kappa_aD[\hat{a}]+\kappa_bD[\hat{b}],
\end{eqnarray}
where $D[\hat{A}]=\hat{A} \hat{\rho} \hat{A}^{\dagger}-\hat{A}^{\dagger}\hat{A}\hat{\rho}/2-\hat{\rho} \hat{A}^{\dagger} \hat{A}/2$ with $\hat{A} = \hat{a},\hat{b}$ or $\hat{c}$ is a Lindblad operator~\cite{Lindblad1976} describing the loss effect of the cavity $A$, cavity $B$ or the magnon, respectively. For convenience, we assume that $\kappa_a=\kappa_b=\gamma_m=0.1g$ that is accessible in experiment~\cite{Huebl2013,Tabuchi2014}. Through solving the master equation with losses, we can obtain time evolution of the fidelity and  collective coherence of $|\phi_1\rangle$ with thick lines in Fig.~\ref{f4}(a), $|\phi_2\rangle$ and $|\phi_3\rangle$ with thick lines in Figs.~\ref{f6}(a) and (b), respectively. The fidelity of $|\phi_1\rangle$, $|\phi_2\rangle$ and $|\phi_3\rangle$ are 0.926, 0.856 and 0.793, respectively. And the collective coherence of $|\phi_1\rangle$, $|\phi_2\rangle$ and $|\phi_3\rangle$ are 0.678, 0.794 and 0.848, respectively. Tripartite high-dimensional entangled state $|\phi_3\rangle$ is most affected by losses due to the largest particle number $N$.

\begin{figure}
\centering
\includegraphics[width=0.9\linewidth]{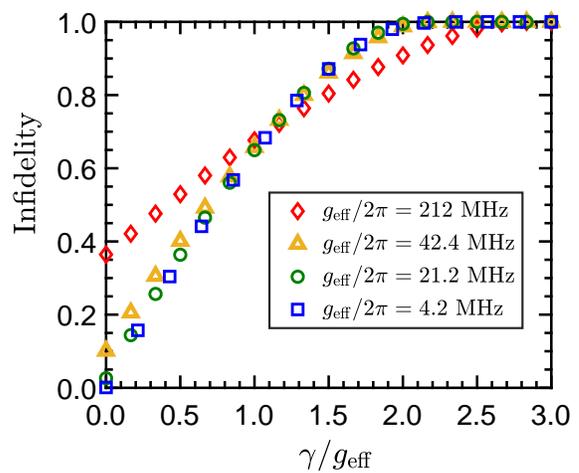}	
\caption{Infidelity of the steady tripartite high-dimensional entangled state versus the decay ratio with different magnon-photon coupling strengths. We choose $\kappa_a=\kappa_b=\gamma_m=\gamma$ and $g_{\rm{eff}}/2\pi=212$ MHz, 42.4 MHz, 21.2 MHz and 4.2 MHz, respectively. Parameters:  $\omega_c/2\pi=6~$GHz,~$\omega_a/2\pi=\omega_b/2\pi=5.95$~GHz,~$g/2\pi=6$~MHz,~$r/2\pi=50$~MHz, $\theta=1.1\pi$ and $T_\mathrm{total}=0.2$~$\mu$s.}\label{f9}
\end{figure}

In Fig.~\ref{f9}, we plot the  infidelity~($1-F$) of tripartite high-dimensional entangled state $|\phi_3\rangle$ at the end of evolution time $T_\mathrm{total}=0.2$~$\mu$s as a function of losses of the cavities and the magnon with different coupling strengths $g_{\rm{eff}}$ in the $\mathcal{PT}$-symmetry broken phase. As the decay rate $\gamma$ increases, the infidelity
of tripartite entangled state shows a trend of increase. When $g_{\rm{eff}}/2\pi=\{42.4,21.2,4.2\}$~MHz, the three lines of infidelities are of insignificant differences, indicating that the magnon-photon coupling strength has little effect on the fidelity as $4.2$~MHz $\leqslant g_{\rm{eff}}/2\pi\leqslant$ $42.4$~MHz. However, under the condition of $g_{\rm{eff}}/2\pi=212$~MHz, the infidelity reaches 0.37 at the point $\gamma/g_{\rm{eff}}=0$, because of the magnon-photon coupling strength mismatching the large detuning condition $ |\omega-\omega_c-r|\gg|g|$. The fidelity of the tripartite high-dimensional entangled state $|\phi_3\rangle$ can be over $90\%$ with $\gamma/g_{\rm{eff}}<0.1$, when the large detuning condition is well satisfied.

\begin{figure}
\centering
\includegraphics[width=0.9\linewidth]{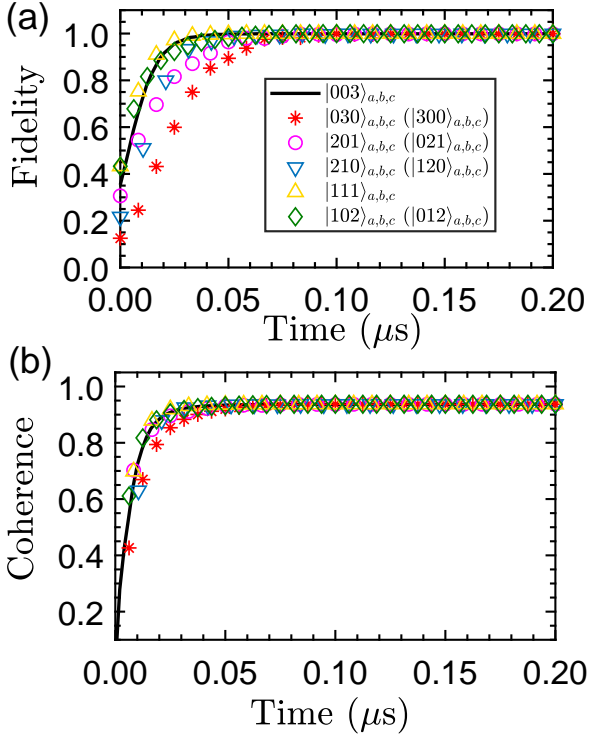}	
\caption{Time dependence of (a) fidelity of the tripartite high-dimensional entangled state $|\phi_3\rangle$ and (b) collective coherence with different initial states by satisfying the condition of the mean particle number equal to $3$. Parameters:  $\omega_c/2\pi=6~$GHz,~$\omega_a/2\pi=\omega_b/2\pi=5.95$~GHz,~$g/2\pi=6$~MHz, $r/2\pi=50$~MHz and $\theta=1.1\pi$.}\label{f10}
\end{figure}

In the protocol above, the tripartite high-dimensional entangled state $|\phi_3\rangle$ is achieved through setting the initial state as $|\psi_3\rangle=|003\rangle_{a,b,c}$. To test dependence on the initial state, we plot the time evolution of the fidelity and collective coherence for $|\phi_3\rangle$ with different initial states satisfying the condition of the mean particle number $N=3$ in Figs.~\ref{f10}(a) and (b), respectively. Interestingly, both of high fidelity and steady collective coherence can be attained by choosing different initial states to meet the condition of $N=3$. In particular, different initial states for the collective coherence have almost the identical evolution trend. Thus, the dynamic evolution of the system is independent of a certain initial state, which can be satisfied with the condition of ${\partial N}/{\partial t}=0$ proved by the commutation relation $[N, \hat{H}] = 0$. According to various of initial states in Fig.~\ref{f10}(a), the shortest evolution time for achieving unity fidelity of $|\phi_3\rangle$ is different, determined by the population of the initial state at $T=0$. Therefore, we can choose the initial state which is of the largest population in the tripartite entangled state (e.g., $|101\rangle_{a,b,c}$, $|002\rangle_{a,b,c}$ and $|011\rangle_{a,b,c}$ for $|\phi_2\rangle$;  $|111\rangle_{a,b,c}$,$|102\rangle_{a,b,c}$ and $|012\rangle_{a,b,c}$ for $|\phi_3\rangle$) so that the target entangled state with the saturation fidelity can be generated more efficiently.

\begin{figure}
\centering
\includegraphics[width=0.9\linewidth]{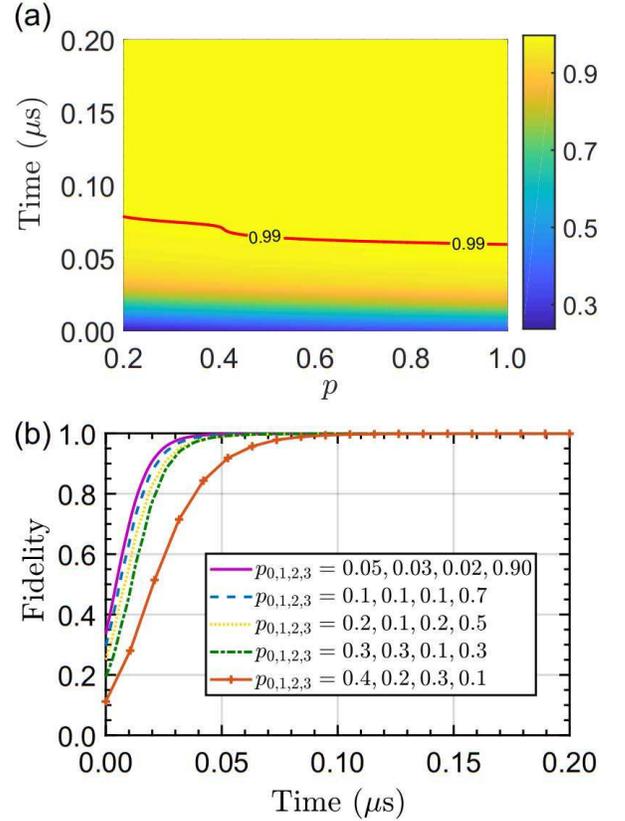}	
\caption{(a) The fidelity of the steady high-dimensional entangled state $|\phi_3\rangle$ versus the evolution time and the purity $p$ of the initial state with respect to $|\tilde03\rangle_{A_1,c}$. The red solid line represents 0.99 fidelity contour line of  the steady high-dimensional entangled state $|\phi_3\rangle$. (b) Time evolution of the fidelity for the steady high-dimensional entangled state $|\phi_3\rangle$ with the initial density: $\rho_0'=\sum_{n=0}^{3}p_n|\tilde{0}n\rangle_{A_1,c} \langle \tilde{0}n|$. Parameters: $\omega_c/2\pi=6~$GHz,~$\omega_a/2\pi=\omega_b/2\pi=5.95$~GHz,~$g/2\pi=6$~MHz,~$r/2\pi=50$~MHz and $\theta=1.1\pi$.} \label{f11}
\end{figure}

Due to influences of external temperature and experimental techniques, the initial state of the hybrid system may not be prepared perfectly as a pure state. Thus, it is convenient to introduce a parameter $p$ that characters the purity of initial state so as to identify the robustness of generating the desired high-dimensional entanglement. The initial state is assumed with a mixed state represented as $\rho_0=p|\tilde{0}3\rangle_{A_1,c} \langle \tilde{0} 3|+(1-p)|\tilde{3}0\rangle_{A_1,c} \langle \tilde{3} 0|$. As shown in Fig.~\ref{f11}(a), we plot the fidelity of $|\phi_3\rangle$ versus the evolution time $T$ and the purity $p$ of initial state with respect to $|\tilde03\rangle_{A_1,c}$. We can learn that the fidelity is not dependent of a certain initial state but the evolution time. The non-Hermitian property of the hybrid system in the $\mathcal{PT}$-symmetry broken phase results in three eigenmodes, whose imaginary parts are positive, negative and zero, respectively. The eigenmode with a positive~(negative) imaginary part is corresponding to a gain~(loss) mode. The particle number of the system in the gain mode will increase until all particles are in this state. That is, the hybrid system behaves as a attractor, while in the $\mathcal{PT}$-symmetry broken phase each mixed state is asymptotically purified to a ground state~\cite{Brody2012}. Specially, we plot a 0.99 fidelity contour line of the steady high-dimensional entangled state $|\phi_3\rangle$ in Fig.~\ref{f11}(a). When the initial state is pure with $p=1$, the shortest evolution time is 0.06~$\mu$s to reach the fidelity $F=0.99$. By comparison with the initial density matrix $\hat\rho_0=0.2|\tilde{0}3\rangle_{A_1,c} \langle \tilde{0} 3|+0.8|\tilde{3}0\rangle_{A_1,c} \langle \tilde{3} 0|$, the steady high-dimensional entangled state with $F=0.99$ requires $T=0.08~\mu$s at least. The red line indicates that a larger proportion of $|\tilde{0}3\rangle_{A_1,c}$ in the initial state demands a shorter evolution time to realize a saturation fidelity of the tripartite high-dimensional entangled state. On the other hand, we further consider an initial state as $\rho_0'=\sum_{n=0}^{3}p_n|\tilde{0}n\rangle_{A_1,c} \langle \tilde{0}n|$ where the total particle number is less than $3$. In Fig.~\ref{f11}(b), the steady entangled state is always of a high fidelity at the end of evolution, showing independence of creating the tripartite high-dimensional entangled state on a certain initial state with the total particle number being either equal to or less than $3$. Furthermore, in Fig.~\ref{f11}(b), we can increase the proportion of the state $|\tilde{0}3\rangle_{A_1,c}$ in the initial state so as to shorten the evolution time to attain the saturation fidelity of $|\phi_3\rangle$, when the initial state is mixed with $|\tilde{0}1\rangle_{A_1,c}$, $|\tilde{0}2\rangle_{A_1,c}$ and $|\tilde{0}3\rangle_{A_1,c}$.

\begin{figure}
\centering
\includegraphics[width=0.90\linewidth]{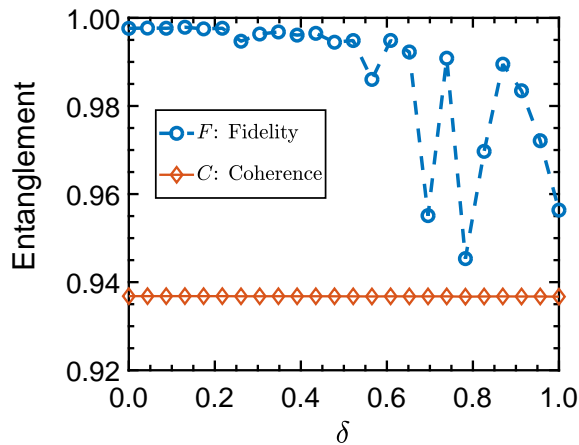}	
\caption{Collective coherence and the fidelity of the steady tripartite high-dimensional entangled state as functions of disorder $\delta$ of the magnon-photon coupling strength $g$. Parameters:  $\omega_c/2\pi=6~$GHz,~$\omega_a/2\pi=\omega_b/2\pi=5.95$~GHz,~$g/2\pi=6$~MHz,~$r/2\pi=50$~MHz, $\theta=1.1\pi$ and $T_\mathrm{total}=0.2$~$\mu$s.}\label{f12}
\end{figure}

In addition, the existence of some circuit imperfections in the magnon-circuit-QED hybrid system, such as the mutual inductance, the self-inductance of circuit, and the unstable external magnetic field, may cause a variation in ideal couplings. In order to study the robustness of the protocol against the variation of cavity-magnon coupling strength, we add a random disorder into coupling strength $g'=g (1+\rm{rand}[-\delta,\delta])$ where  $\rm{rand}[-\delta,\delta]$ is to pick up a random number in the range of $[-\delta, \delta]$. The relation among the fidelity of steady entangled state $|\phi_3\rangle$, collective coherence and the disorder $\delta \in [0,1]$ is exhibited in Fig.~\ref{f12}. It is the disorder that is randomly sampled 51 times, and then the fidelity and collective coherence are taken as an average of the 51 results. Learning from the red-diamond line unchanged with varying $\delta$, the collective coherence is independent of the disorder of cavity-magnon coupling strength, which reflects the robustness due to the non-Hermitian property of  $\mathcal{PT}$-symmetry broken phases.  Under the condition of $\delta \in [0,0.5]$, the fidelity always remains above $99.5\%$. Nevertheless, the fidelity oscillates obviously between $94.5\%$ and $99.2\%$ with $\delta \in [0.65,1]$, on account of fluctuations of parameter $g$ affecting approximate conditions to produce an effective Hamiltonian~(\ref{e8}) for the steady tripartite high-dimensional entangled state.

\section{Conclusion}\label{S7}
To summary, we have proposed a non-Hermitian model of the magnon-circuit-QED hybrid system. There are the steady quantum coherence and tripartite high-dimensional entangled states among the modes of magnon and photon in $\mathcal{PT}$-symmetry broken phases in the proposed system. The tripartite high-dimensional entangled state and the quantum coherence are robust to the dissipation of hybrid system and the fluctuation of magnon-photon coupling. Besides, the tripartite high-dimensional entangled state is independent of a certain initial state. We take into account the experimental considerations, including the implementation of the model, the design of equivalent circuit diagram and the realization of non-Hermitian coupling between the modes of magnon and photon in the circuit. This work provides a new approach to generate tripartite high-dimensional entangled states and is expected to be helpful for realizing tripartite and even multipartite high-dimensional entangled states in the non-Hermitian system with the hybridization of the magnon and the circuit-QED system.
\section*{ACKNOWLEDGEMENTS}
This work was supported by National Natural Science Foundation of China (NSFC) (Grant No.11675046), Program for Innovation Research of Science in Harbin Institute of Technology (Grant No. A201412), and Postdoctoral Scientific Research Developmental Fund of Heilongjiang Province (Grant No. LBH-Q15060).
\bibliography{apssamp}
\end{document}